\documentclass{aa}  
\usepackage{graphicx}
\usepackage{txfonts}
\usepackage{natbib}
\bibpunct{(}{)}{;}{a}{}{,}
\begin{document}
   \title{A Study of Catalogued Nearby Galaxy Clusters in the SDSS-DR4.}

   \subtitle{II. Cluster Substructure}

   \author{J. Alfonso L. Aguerri \and Rub\'en S\'anchez-Janssen}

   \offprints{J. A. L. Aguerri}

   \institute{Instituto de Astrof\'{\i}sica de Canarias. C/ V\'{\i}a L\'actea s/n, 38200 La Laguna, Spain\\
              \email{jalfonso@iac.es, ruben@iac.es}
             }

   \date{Received ; accepted }

 
  \abstract
   {According to the current cosmological paradigm, large scale structures form hierarchically in the Universe. 
   Clusters of galaxies  grow through a continuous accretion of mass, and the presence of cluster substructures can be interpreted as signature of this process. Nevertheless, the rate 
   and manner of  mass accretion events are still matters of debate.}
   {We have analysed the presence of substructures in one of the largest sample of nearby cluster galaxies available 
   in the literature. We have determined the fraction of clusters with substructure and the properties of the galaxies located in such substructures.}
   {Substructure in the galaxy clusters was studied  using the Dressler--Shectman test, which was calibrated through 
   extensive Monte Carlo simulations of galaxy clusters similar to  real ones. In order to avoid possible 
   biases in the results due to differing incompleteness among clusters, we  selected two galaxy populations: a) 
   galaxies brighter than M$_{r} = $-20 located in clusters at $z < 0.1$ (EC1); and b) galaxies of brightness $M_{r} < -19$ located at $z<0.07$ (EC2).}
   {In the inner cluster regions ($r < r_{200}$) 11$\%$ and 33$\%$ of the clusters of EC1 and EC2 respectively show substructure. 
   This fraction is larger in the outer cluster regions ($\approx 55\%$) for EC1 and EC2 samples. Cluster global properties, such as $\sigma_{c}$, $f_{b}$ or $\Delta m_{12}$ do not depend on the amount of cluster substructure. We have studied the properties of individual galaxies located in substructures in the EC1 and EC2 galaxy 
   populations. The fraction of galaxies within substructures is larger in the outer cluster regions when fainter galaxies 
   are included. The distribution of relative velocities of galaxies within substructures suggest that they consist of an infalling population mixed with backsplash galaxies. We can not rule out that the infalling galaxy population located in substructures are genuine field ones.}
   {}


   \maketitle
%

\section{Introduction}

Pioneering photographic surveys \citep[e.g.][]{abell58, zwicky68, shectman85} showed that galaxies were distributed not uniformly 
but in clumps, forming  aggregations called galaxy clusters. 
These gravitationally bound systems are (together with galaxy superclusters) the largest and most massive  structures 
known in the Universe. These large galaxy aggregations have been formed by merging smaller galaxy units \citep[e.g.][]{white78}. 

The increasing availability of galaxy surveys during the last decade has shown that a considerable fraction of nearby 
galaxy clusters are not yet virialized systems  \citep[see e.g.][and references therein]{ramella07}. The most obvious observational signature is the presence of multi-modality 
in the spatial and velocity distributions of galaxies and/or gas. Substructure can be defined as the presence of two or more galaxy aggregations within a cluster of galaxies. These galaxy aggregations could be associated with late cluster formation due to a recent merger 
of small groups forming a cluster,  secondary infalls of galaxy groups in a virialized cluster, or gravitationally bound 
subclusters still in an infalling process. Studying cluster substructure 
therefore allows us to investigate the process by which clusters form. Indeed, substructure in galaxy clusters depends on the expansion rate of the Universe, and  its study can also constrain the cosmological model for structure formation and our understanding of 
the nature of dark matter \citep{richstone92, kauffmann93, lacey93, mohr95, thomas98, markevitch04, clowe06}. Moreover, the 
identification of cluster substructure is an important issue, since they can strongly affect estimates of global cluster parameters such as mass and 
velocity dispersion \citep{perea90, escalera94, girardi96}.

The presence of substructure in clusters can also influence the evolution of the constituent galaxies. Galaxy clusters are popular sites 
for studying environmental influences on galaxy evolution. Cluster galaxies are substantially different from galaxies in the 
field. This idea is supported by many observational results, such as the morphology-density relation \cite[][]{dressler80}, the low 
fraction of star-forming galaxies \cite[e.g.][]{popesso07}, or the different structural parameters  shown by galaxies in high density environments 
with respect to the field \cite[][]{aguerri04, gutierrez04, barazza09, marinova09, aguerri09a, aguerri09b}. It was previously thought that cluster and field galaxies
 had different origins. Nowadays, these differences are explained in terms of the different evolution of the high density and the field 
 galaxy population. Hence, all previous observational results could be explained in terms of several large- and small-scale mechanisms present in 
 high density environments and not in the field, among which we can mention galaxy harassment \cite[][]{moore96}, ram-pressure stripping 
 \cite[][]{gun72, quilis01}, tidal stripping \cite[][]{merritt83, merritt84}, major and minor galaxy mergers \cite[][]{aguerri01, elichemoral06}, 
 and starvation \cite[][]{larson80}. Within this framework of galaxy evolution, where the environemnt plays an important role, 
 substructure should be taken into account in seeking a global picture of the evolution of galaxies. There is thus
 observational evidence that a higher fraction of cluster galaxies with ongoing starburst episodes are located in substructures
  or in the regions of cluster--subcluster interactions \citep[][]{caldwell93, biviano97, moss00, poggianti04}. Moreover, it has been observed 
  that the presence of substructures is also directly related to the growth of the brightest cluster galaxies (BCGs). Thus, brighter 
  BCGs are located in clusters with less substructure \citep{ramella07}.

The analysis of substructure can be performed using the projected phase-space distribution of cluster galaxies \citep[e.g.][]{dressler88}, the surface-brightness 
distribution and temperature of the X-ray emitting intra-cluster gas \cite{briel92}, or the shear pattern in the background galaxy 
distribution induced by gravitational lensing \cite{abdelsalam98}. In this paper we focus on the study of the phase-space distribution 
of galaxies in clusters, traditionally studied using different statistical tests. Depending on the information used, those statistical tests can 
be classified as 1D, 2D or 3D. The 1D tests use only the velocity information. The null hypothesis is that the velocity distribution 
in relaxed galaxy clusters should be Gaussian. Thus, these tests study the degree of non-gaussianity of the velocity distribution of 
galaxies analysing (among other things) its skewness, kurtosis, asymmetry index or tidal index \cite{beers90}. The 2D tests assume that the 
galaxy distribution in virialized clusters should follow a Poissonian distribution. The most popular 2D tests are, among others: the angular separation test \cite{west95}, the $\beta$-test \cite{west88}, the Fourier elongation test \cite{mohr93} and the Lee statistic \cite{fitchett87, fitchett88}.  The 3D tests use the  spatial  plus the radial velocity information. The null hypothesis of these tests is that no 
correlation between position and velocity exists. The most popular of these tests are the Lee 3D test, the $\Delta$-test and the $\alpha$-test \cite{west90}. 
Numerical simulations have shown that 3D tests are the most efficient for detecting substructure \cite{knebe00, pinkney96}.

The previously mentioned tests have been applied to different cluster samples and have revealed that about 1/3 of the nearby galaxy clusters are 
not yet virialized systems \cite{west90, rhee91, bird94, escalera94, west95, girardi97, solanes99, biviano02, burgett04, flin06, ramella07}. 
Most of these studies were done on small galaxy cluster samples and/or they contain no velocity information on the galaxies. Few nearby galaxy 
cluster  surveys have  large datasets of redshift information (see e.g. ENACS Biviano et al. 2004; or 2dFGRS cluster sample Burgett et al. 2004) and the redshift information is even more limited for 
galaxy clusters at high redshift. Nevertheless, recent surveys of galaxy clusters at high redshift hint that the fraction of clusters with 
substructure is larger at higher redshift, thereby suggesting that the hierarchical growth of clusters was more intense in the past 
\cite{vandokkum00, haines01, huo04, demarco05, halliday05, milvang08}.

In this paper we have analysed the presence of substructure in one of the largest nearby galaxy clusters sample available in the literature 
\cite{aguerri07}. Substructure has been measured by using the 3D Dressler--Schectman (DS) test which, according to numerical 
simulations, is one of the most efficient tests for detecting substructure in clusters \citep[see][]{knebe00, pinkney96}. Special attention was 
given to the influence on the results of possible observational biases, such as different aperture observations and the comparison of different 
galaxy populations among clusters. We were also interested in the study of galaxy substructure in a broad range of galaxy densities. 
Thus, one of the novel results presented here is the study of substructure in the outskirts $(r<2r_{200})$ of galaxy clusters, a region that has not been
well analysed in previous large cluster samples.

A brief summary of the data is present is Section 2. The global substructure is analysed in Section 3. 
The properties of the galaxies located outside and inside substructures are presented in Section 4. We present our discussion and conclusions 
in Section 5 and 6, respectively. 


\section{Cluster data}

The data used in this study comprises the cluster sample analysed in Aguerri et al. (2007; hereafter ASM07). This sample consists of 88 nearby ($z<0.1$) and 
isolated galaxy clusters with known redshift  catalogued by Abell et al. (1989), Zwicky \& Humason (1961), B{\"o}hringer et al. (2000), or Voges et al. (1999), and mapped by the SDSS-DR4 
(York et al. 2000; Adelman-McCarthy et al. 2006). We only downloaded galaxies located within a radius of 4.5 Mpc around the centres of each 
galaxy cluster. Only clusters with a minimum of 30 galaxies with spectroscopic data in the search radius were considered. In order to avoid 
possible incompleteness effects of the SDSS-DR4 spectroscopic data, we completed the spectroscopic SDSS-DR4 observations with the data available 
in the NASA Extragalactic Database (NED). Thus, we obtained a constant completeness (about 85$\%$) for all magnitudes down to $m_{r}=17.77 $ mag  
(see ASM07 for more details). 

Cluster membership in our sample was determined using a combination of the ZHG algorithm \cite{zabludoff90} and the KMM 
algorithm \cite{ashman94}. A first rough cluster membership determination was obtained using the ZHG algorithm. This is a typical gapping 
procedure that determines cluster membership by the exclusion of those galaxies located at more than a certain velocity distance ($\Delta$v) 
from the nearest galaxy in velocity space. We used $\Delta$v=500 km s$^{-1}$ for our cluster sample (see ASM07 for more details). In the second 
step, cluster membership was refined using the KMM algorithm, which estimates the statistical significance of bi-modality in a dataset. We 
ran the KMM algorithm on each group of galaxies determined by the ZHG algorithm to be galaxy cluster. The KMM algorithm gave us the compatibility of the velocity distribution of such group of galaxies with a single or multiple Gaussian distributions. We considered three different cases: a) Single clusters: the velocity distribution of the galaxies is compatible with a single Gaussian; b) Cluster with substructure: the velocity distribution show multiple groups. We defined the cluster as the most populated group plus those groups whose mean velocities lie within 3$\sigma$ from he mean velocity of the largest one; c) clusters with contaminants: the velocity distribution is compatible with the presence of several groups, but the mean velocities of the smaller groups deviate more than 3$\sigma$ from the most populated one, which we identify as the cluster itself.  It is important 
to note that this cluster membership selection will affect the detected substructure of these clusters. We should keep this procedure in mind when comparing substructure results from different studies. The final total number of galaxies located in clusters was 10865 
objects (see ASM07).   

The 88 galaxy clusters are located in a redshift range between 0.02 and 0.1, with an average redshift of 0.07. This redshift range and the 
fact that the observational data are complete until $m_{r}=17.77$ mean that the completeness absolute r-band magnitude ($M_{r}$) is a function
 of redshift (see Fig. 5 from ASM07). Thus, our sample is complete for galaxies brighter than $M_{r}=-20.0$. It is also important to note that 
 the galaxy data were downloaded from SDSS-DR4 archive according to a metric criterion, and therefore we could be mapping different physical 
 regions for each cluster. Nevertheless, all the clusters in our sample map the region $r<2r_{200}$ (see ASM07), so only galaxies within this radius have been considered in the following substructure analysis. This provides us with a total number of 6880 galaxies located in comparable
  physical regions for each of our clusters.

It is important to considered that our cluster membership is not free of projection effects. Thus, some galaxies could be located at larger physical radii than the projected ones. Rines et al. (2005) showed that $\approx 40\%$ of star-forming and $15\%$ of non-star-forming galaxies located at projected radii between 1-2$r_{200}$ were located at larger physical distances.

   \begin{table}
      \caption[]{Percentage of clusters with substructure}
         \label{KapSou}
     $$ \centering
         \begin{tabular}{cccc}
            \hline\hline
            Radius &  M$_{r}<-20$ & M$_{r}<-19$ & No mag. limit\\
            \hline
           r$<$r$_{200}$ & 11& 33 & 34\\
           r$_{200}<$r$<$2r$_{200}$ & 55& 57 &44\\
            \hline
         \end{tabular}
     $$ 
   \end{table}

\section{Global substructure in clusters}

The cluster substructure was measured using the kinematic and/or spatial information of the galaxies within the clusters. 
In order to account for this, several statistical tests have been developed during last decades. In particular, the method developed by Dressler \& Shectman (1988; hereafter DS) is one of the most 
efficient (Pinkney et al. 1996). The algorithm starts by calculating the mean velocity ($v_{local}$) and standard deviation 
($\sigma_{local}$) for each galaxy of the cluster and its $N_{local}$ nearest neighbours. These local values 
 are compared with the mean velocity $v_{c}$, and standard deviation $\sigma_{c}$ of the global cluster. The deviation of the local from the global kinematics for each galaxy is then defined by:

\begin{equation}
\delta_{i}^{2}=(\frac{N_{local}+1}{\sigma_{c}^2})[(v_{local}-v_{c})^2+(\sigma_{local}-\sigma_{c})^2]
\end{equation}

Finally, a cumulative quantity $\Delta=\sum \delta_{i}$ is computed that serves as the statistic 
for quantifying the substructure. This test is normalized with Monte Carlo simulations in which the velocities 
are shuffled among the positions. In this way, an existing  local correlation between velocities and 
positions is destroyed. The probability of the null hypothesis that there are no such correlations is 
given in terms of the fraction of simulated clusters for which the cumulative deviation is smaller than 
the observed value. We have normalized the statistic of the test with 1000 Monte Carlo simulations per cluster.

Biviano et al. (2002) slightly redefined this test. The parameter $\delta$ was calculated by:

\begin{equation}
\delta_{i}=\frac{1}{\sigma_{c}}\sqrt{\frac{N_{local}\delta_{v}^{2}}{(t_{n_{loc}}-1)^{2}}+\frac{\delta_{\sigma}^{2}}{(1-\sqrt{(N_{local}-1)/\chi^{+}_{N_{local}}-1})^{2}}}
\end{equation}

with $\delta_{v}=|v_{local}-v_{c}|$ and $\delta_{\sigma}=max(\sigma_{c}-\sigma_{local},0)$, 
where the Student-t and $\chi^{2}$ distributions were used to calculate the uncertainty in the velocity 
and velocity-dispersion differences, respectively. This definition of the $\delta$ was designed to obtain 
groups of galaxies that are colder than the cluster and/or have an average velocity that differs from the 
global cluster mean. As before,  $\Delta=\sum \delta_{i}$. This was the version of the test 
used in our study. Nevertheless, we ran the test on our data weighting or not weighting $\delta_{v}$ 
and $\delta_{\sigma}$ and obtained similar results. 

The results produced by the test depend on the value of N$_{local}$ and the number of galaxies per cluster. 
Originally, Dressler \& Shectman (1988) proposed the computation of $\Delta$ using $N_{local}=10$ 
independently of the number of galaxy cluster members. Pinkney et al. (1996) demonstrated that 
 the detection of substructure was more efficient when $N_{local}$ depends on the number of cluster members. 
They proposed $N_{local}=\sqrt{N_{gal}}$, $N_{gal}$ being  the number of cluster members. Numerical simulations 
also show that the DS test effectively detects substructures in clusters with more than 30 galaxies. 
Here, we have studied the substructure of those galaxy clusters with $N_{gal}> 30$ by using $N_{local}=\sqrt{N_{gal}}$. Biviano et al. (2002) computed $\delta_{i}$ for each galaxy as the average of the $\delta$-values of its N$_{local}$-1 neighbours. We did not find any important differences when carrying out this smoothing, so we finally decided to work with the unsmoothed $\delta$-values of each galaxy.

\subsection{Testing the substructure code}

In order to test the code used for detecting substructure in our galaxy clusters we ran several
 Monte Carlo simulations of clusters. These simulated clusters did not hold substructure and were used in 
order to test the false positives cases detected by our code, i.e. clusters with substructure reported by 
the test but without substructure. We considered all our clusters with more than 30 galaxies 
in $r<2 r_{200}$, and substituted their velocity distribution by a Gaussian distribution with the 
same v$_{c}$ and $\sigma_{c}$ as the real clusters. This kind of simulation breaks the possible correlation 
between galaxy position and velocity, i.e. possible substructure. We ran the DS test on these simulated 
clusters as for the real ones, giving no signal of substructure in all cases. This means that our test does not
 detect cases of false positives in the study of the global substructure in clusters.

We have also ran another set of MC simulations of clusters in order to analyse the efficiency of the DS test 
for detecting substructure. In contrast to the previous set of simulations, these simulated clusters 
held substructure. Thus, the new simulated clusters had two populations of galaxies: one with no 
correlation between position and velocities showing a Gaussian velocity distribution (we call this a relaxed 
population) and another of galaxies showing substructure (we call this an unrelaxed population). 
The surface distribution of the galaxies of the relaxed and unrelaxed components was simulated following a 
Hernquist distribution \cite[see][]{hernquist90, mahdavi99, rines06}. The relaxed galaxies were located within  a radius 
of r/r$_{200} \leq 2$, while the unrelaxed component represents a compact galaxy group with radius r/r$_{200}\leq0.2$, 
being the centre of the unrelaxed component randomly chosen. The velocity distribution of the relaxed 
and unrelaxed components is Gaussian with different mean and sigma. We have selected 7 
different values per simulated cluster for the relative velocity of cluster and group: 
$(v_{c}-v_{g})/\sigma_{c}$= 0, 0.5, 1.0, 1.5,2.0, 2.5, and 3.0, v$_{c}$ and $\sigma_{c}$ being the mean velocity 
and velocity dispersion of the relaxed component, and v$_{g}$ the mean velocity of the unrelaxed galaxy population. 
We also varied the ratios between the velocity dispersion of the relaxed and unrelaxed components, choosing 
$\sigma_{g}/\sigma_{c}=0.3, 0.5$ and 0.8 for our simulations. We simulated two sets of clusters with different 
numbers of galaxies (50 and 100). The unrelaxed component in each of these clusters represents 10$\%$ and 
30$\%$ of the total galaxy population respectively. Thus, for a fixed total number of galaxies we have simulated 42 different 
clusters varying $v_{g}-v_{c}$, $\sigma_{g}/\sigma_{c}$ and the percentage of substructure. Each of these simulated 
clusters was simulated 100 times changing the centre of the unrelaxed component. In total, 8\,400 different clusters were simulated.

\begin{figure*}
   \centering
   \includegraphics[width=15cm]{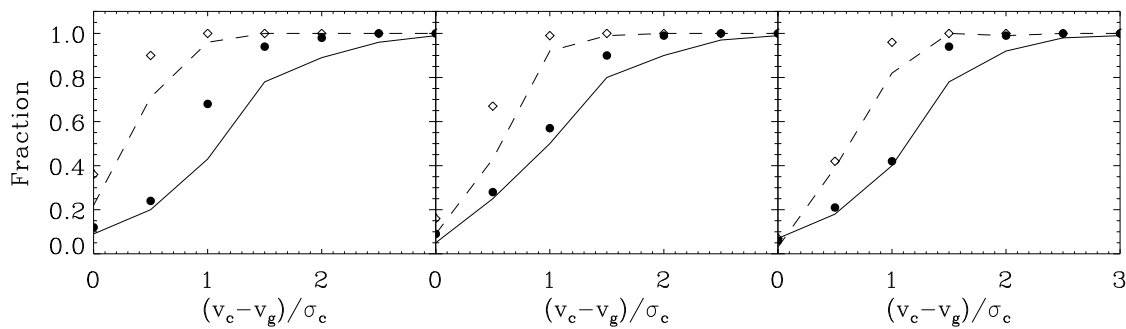}
      \caption{Fraction of detected clusters with global substructure as a function of $(v_{c}-v_{g})/\sigma_{c}$ 
for MC simulated clusters with $\sigma_{c}/\sigma_{g}=$0.3 (left panel), 0.5 (central panel), and 0.7 (right panel). 
In all panels the symbols represent clusters with a population of relaxed:unrelaxed component given by: 90:10 
(full points), 70:30 (diamonds), 45:5 (full line) and 35:15 (dashed line). See text for more details.
              }
         \label{f1}
   \end{figure*}

We computed the fraction of simulated clusters with substructure  provided by the DS test. 
Figure \ref{f1} shows this fraction as a function of $(v_{c}-v_{g})/\sigma_{c}$. Note that the DS test is not 
very efficient for detecting substructure in clusters when the relaxed and unrelaxed components have similar 
mean velocities. In contrast, the efficiency is very high (more than 80$\%$) for clusters with 
$(v_{c}-v_{g})/\sigma_{c}>1-1.5$. Note also that, for a given value of $(v_{c}-v_{g})/\sigma_{c}$, the substructure 
is detected more eficiently when it is located in cooler groups, and represents a larger fraction of the galaxies in the cluster. These constraints were also pointed out by Pinkney et al. (1996) 
in their numerical simulations. 

\subsection{Results on the global substructure} 

We applied the DS test to our galaxy cluster sample. We analysed the substructure of the clusters,
taking into account the galaxies inside two apertures: $r<r_{200}$ (inner cluster regions) and 
$r_{200}<r<2r_{200}$ (outer cluster regions). In the inner cluster regions, 34$\%$ of our 
galaxy clusters showed substructure at the 95$\%$ significance level . This value is compatible with 
other results obtained previously using the same statistics. Thus, Solanes et al. (1999), Biviano et al. (1997), 
Escalera et al. (1994), Bird (1994), and Biviano et al. (2002) found substructure in 
31$\%$ (21/67), 40$\%$, 38$\%$ (6/16), 44$\%$ (11/25), and 40$\%$ (9/23) of their clusters respectively. 
In contrast, Dressler \& Shectman (1988) found a larger percentage of clusters with substructure: 53$\%$ (8/15). 
The amount of substructure in the outskirts of our clusters is larger. In this case, 44$\%$ of the clusters 
show substructure at the  95$\%$ significancd level.

\subsection{Substructure and galaxy population}

The value of $\Delta$ was computed taking into account the local environment of each galaxy in the clusters. 
Nevertheless, different galaxy populations could show different substructure properties. Therefore, one individual cluster 
could show substructure --or not-- depending on the galaxy population used to computed $\Delta$. Thus, the study 
of the substructure in clusters could be biased if the clusters in the  sample does not have a uniform absolute magnitude completeness.

Our galaxy cluster sample does not show a uniform absolute magnitude completeness (see Fig. 5 in ASM07).  All the
clusters have data for the galaxy population brighter than $M_{r}=-20.0$, but fainter galaxies are lost as the 
redshift increases. Therefore, different clusters trace different galaxy populations. In order to analyse the
influence of this on the study of the substructure we ran the DS test on two galaxy population samples. The first 
consisted of those galaxies brighter than $M_{r}=-20.0$ located in clusters at $z<0.1$. The second galaxy 
population is formed by those galaxies brighter than $M_{r}=-19.0$ located in the clusters at $z<0.07$. In accordance with
fig. 5 of ASM07, these restrictions allow us to have two families of clusters with uniform galaxy populations.

Table 1 shows the percentage of clusters with substructure in their inner and outer regions, taking into account 
the previous restrictions on redshift and galaxy absolute magnitudes. We can see that, independently of the galaxy 
population, the percentage of clusters with substructure is higher when the galaxies in their outskirts are considered. 
In the inner cluster regions ($r<r_{200}$) few clusters ($11\%$) show substructure if we consider galaxies 
brighter than $M_{r}=-20.0$. That percentage increases to 33$\%$ when a fainter galaxy population is considered. 
This indicates that the detected fraction of clusters with substructure in the inner regions strongly depends on the 
limiting magnitude of the surveys. Table 1  also shows that the percentage of clusters with substructure does not 
depend strongly on the galaxy population in the outer cluster regions.

   \begin{table*}
      \caption[]{Properties of clusters with and without substructure}
         \label{KapSou}
     $$ \centering
         \begin{tabular}{ccccc}
            \hline\hline

                Mean  &   \multicolumn{2}{c}{M$_{r}<-20$} &  \multicolumn{2}{c}{M$_{r}<-19$}  \\
                   & Sub & No Sub & Sub & No Sub \\
            \hline
     $\sigma_{c}$ (km s$^{-1}$) & 588.5$\pm$199.5 & 544.8$\pm$199.2 & 567.4$\pm$250.7 & 512.5$\pm$150.3 \\
     $f_{b}$ & 0.37$\pm$0.06 & 0.42$\pm$0.12 & 0.36$\pm$0.10 & 0.36$\pm$0.08 \\
     $\Delta m_{12} (mag)$ & -0.45$\pm$0.5 & -0.38$\pm$0.4 & -0.39$\pm$0.47 & -0.36$\pm$0.34 \\
     \hline
         \end{tabular}
     $$ 
   \end{table*}

\subsection{Substructure and global cluster properties}

We have analysed some global properties of the clusters with and without substructure, such as the fraction of 
blue galaxies ($f_{b}$), cluster velocity dispersion ($\sigma_{c}$) and the luminosity difference between the 2 
brightest cluster galaxies ($\Delta m_{12}$). For this comparison we have considered galaxies within an aperture 
$r<r_{200}$. In the previous subsection, we have seen that mapping different galaxy populations in clusters could 
affect the percentage of clusters with substructure. We should therefore take this into account and select a cluster 
sample with a galaxy population free of this bias. We have also selected clusters with more than 30 galaxies. This makes a final number of 28 and 32 clusters when we consider galaxies with M$_{r}<-20$ and M$_{r}<-19$, respectively.

\begin{figure*}
   \centering
   \includegraphics[width=15cm]{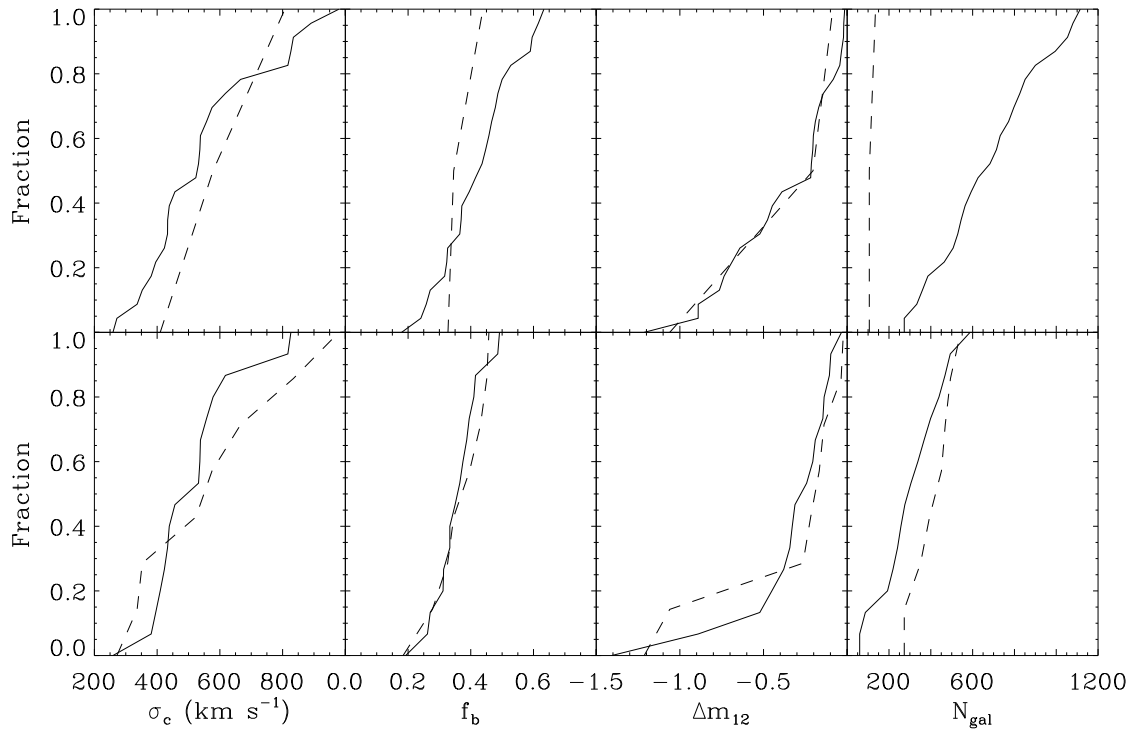}
      \caption{Cumulative distribution funcstions of $sigma_{c}$,$f_{b}$, $\Delta m_{12}$, and galaxy members for clusters with (dashed line) and without (full line) substructure. Two populations of galaxies were considered: brighter than $M_{r}=-20$ (top panels) and brighter than $M_{r}=-19.0$ (bottom panels).
              }
         \label{subes}
   \end{figure*}

Table 2 shows the mean values of $\sigma_{c}$, $f_{b}$ and 
$\Delta m_{12}$ for clusters with and without substructure. These values were obtained taking into account galaxies 
in clusters at $z<0.1$ and brighter than M$_{r}=-20$, and those located in clusters at $z<0.07$ and brighter than
 M$_{r}=-19$. No differences in the clusters properties were observed in the two samples of 
 galaxies (see Tab. 2). Figure \ref{subes} shows the cumulative distribution functions of $\sigma_{c}$, $f_{b}$ and $\Delta m_{12}$ for cluster with and without substructure. This figure also shows the cumulative distribution functions of the number of galaxy members in clusters with and without substructure. Notice that the number of clusters with substructure is very small when galaxies brighter than $M_{r}=-20.0$ are considered. However, this number is larger when we study galaxies brighter than $M_{r}=-19.0$. In this case, the Kolmogorov-Smirnoff (KS) test does not report statistical differences in the cluster properties for clusters with and without substructure. Thus, substructure in the inner cluster region does not affect $\sigma_{c}$, $f_{b}$ or 
$\Delta m_{12}$. 

The above result is not in agreement with Ramella et al. (2007). They found a clear difference between the mean value of $\Delta m_{12}$ for clusters with and without substructure. The work by Ramella et al. was based on a sample of 77 nearby clusters (0.04$<z<$0.07) from the WINGS survey (Fasano et al. 2006). The substructure was determined using the DEDICA procedure (Pisani 1993, 1996). This procedure was aplied to all galaxies in the cluster sample brighter than $M_{V}=-16$. The different galaxy population studied could be the reason of the disagreement with Ramella et al. (2007).

%

\section{Galaxies in substructure}

Instead of summing all individual values of $\delta$ to get an estimate of the amount of substructure in the 
clusters as a whole, we can use the values of $\delta$ for each galaxy in order to select those galaxies located 
in substructures. The problem is to determine the value $\delta_{c}$, which effectively separates galaxies 
inside and outside substructures. We have taken into account two different approaches to this problem. 
The first approach is to consider different values of $\delta_{i}$ for each cluster. In 
contrast, we can also take a unique value of $\delta_{g}$ for all clusters. We tested both methods in order 
to find out which one was optimal.

We selected the values of $\delta_{i}$ for each cluster by comparing the cumulative distribution 
functions of the values of $\delta$ for the galaxies and those from the Monte Carlo simulations of each cluster 
used for normalizing the DS test. These simulations broke the correlation between position and velocity of 
the galaxies. Therefore, they should provide the expected values of $\delta$ of galaxies not located in substructures. 
Three different values were selected for each cluster: the maximum value of $\delta$ obtained from the MC simulations 
($\delta_{i,max}$) and the values where the cumulative distribution function of $\delta$ values of the simulations 
was equal to 0.99 ($\delta_{i,99}$) and 0.999 ($\delta_{i,99.9}$). 

The number of galaxies in individual samples is sometimes too small for segregation studies. Therefore, the 
combination of data from many clusters will make the statistical analysis more reliable. Indeed, the determination of galaxies 
in substructure by adopting a unique value of $\delta_{g}$ for all clusters is based on the combination of data in an 
ensemble cluster. We built an ensemble cluster by normalizing the scales and velocities for each galaxy. The 
radial distance of each galaxy to the cluster centre was therefore scaled by the $r_{200}$ value for the corresponding cluster, and 
the relative velocity of each cluster galaxy was normalized by the velocity dispersion of the cluster. As pointed out by 
Biviano et al. (2002), the ensemble cluster is made on the implicit assumption that the distributions of galaxy 
types are similar in the individual clusters. Nevertheless, those distributions can be different for several reasons,
such as  different observing apertures of the clusters and different galaxy incompleteness. 
Our ensemble cluster is free from aperture bias because all our clusters are contributing at all radial distances up to 2$r_{200}$. But we are not free 
from possible biases due to galaxy incompleteness (see Fig. 5 in ASM07). This bias was avoided by building 
two ensemble clusters. The first was built by those galaxies brighter than $M_{r}=-20$ located in 
clusters at $z<0.1$ (hereafter EC1), and the second ensemble cluster was formed with galaxies brighter than 
$M_{r}=-19.0$ in clusters at $z<0.07$ (hereafter EC2). In both cases only clusters with more than 30 galaxies
 within 2$r_{200}$ contribute to the ensemble clusters. Taking into account all previous restrictions,
 EC1 was built with 2593 galaxies from 44 galaxy clusters, and EC2 was formed by 2400 galaxies from 34 clusters.

A global value for $\delta_{g}$ was determined by analysing the cumulative distribution function of 
the $\delta$ values of the Monte Carlo simulations used in the normalization of the DS test. We selected 
two values of $\delta_{g}$ for which the cumulative $\delta$ distribution function of the MC simulations 
was equal to 0.95 ($\delta_{g,95}$) and 0.99 ($\delta_{g,99}$). For our two ensemble clusters these values turned to be
 $\delta_{g,95}=$1.98 and 2.03 and $\delta_{g,99}=$2.7 and 3.0 for EC1 and EC2, respectively. Galaxies with 
$\delta$ values larger than these global values were considered as galaxies in substructures. Note that our 
$\delta_{g,95}$ value is similar to that adopted by Biviano et al. (2002).

\begin{figure*}
   \centering
   \includegraphics[width=15cm]{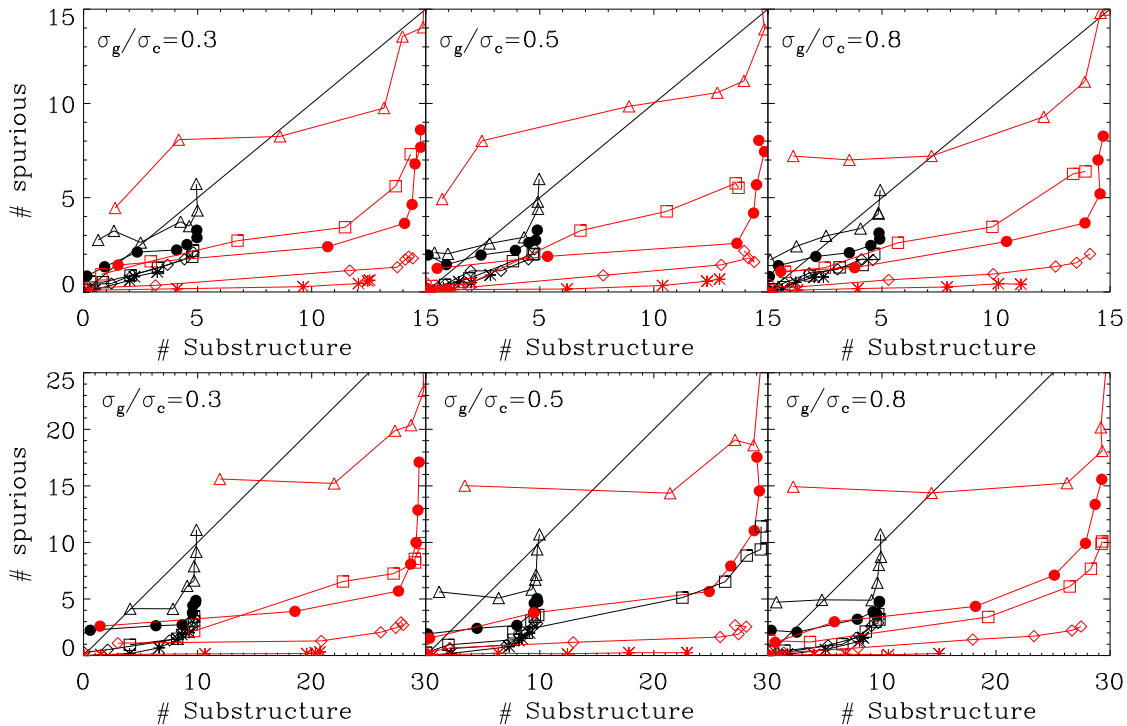}
      \caption{The number of galaxies detected in substructure versus spurious galaxies detected as substructure 
in the Monte Carlo simulated galaxy clusters. The values of $\delta_{c}$ used for identifying galaxies in 
substructure were: $\delta_{g,95}$ (triangles), $\delta_{g,99}$ (squares), $\delta_{i,99}$ (points), 
$\delta_{i,99.9}$ (diamonds) and $\delta_{i,max}$ (asterisks). The upper and lower panels represent clusters containing  50 and 100 galaxies respectively. The black and red symbols correspond to clusters with 10$\%$ and 
30$\%$ of galaxies in substructure respectively. From left to right, the symbols in all the panels represent 
clusters with $(v_{c}-v_{g})/\sigma_{c}$=0, 0.5, 1.0, 1.5, 2.0, 1.5 and 3.0
              }
         \label{f2}
   \end{figure*}

We have used the MC simulated clusters reported in Sec. 3.1 for determining the best value of $\delta_{c}$ for 
detecting galaxies in substructures. The substructure in the simulated clusters were analysed by the DS test 
following the same prescription as the real ones. In the simulated MC clusters we know whether the galaxies selected 
as substructure by the DS test are real or spurious substructure galaxies. It is expected that the number of 
galaxies in substructure will strongly depend on the adopted value of $\delta_{c}$. The best $\delta_{c}$ values should 
maximize the detection of galaxies in substructure and minimize the spurious ones. Figure ~\ref{f2} shows the number 
of galaxies detected as substructure versus the number of spurious substructure detections for the simulated galaxy 
clusters and for the different values of $\delta_{c}$. The results from simulated clusters with 50 and 100 galaxies containing 10$\%$ or 30$\%$ of their population in substructures are shown in Fig. ~\ref{f2}. Similarly to what was reported in Sec. 3.1, the DS test 
does not detect individual galaxies in substructures when the relaxed and the unrelaxed
  component have similar mean velocities.  It is also evident from Fig. ~\ref{f2} that studies of substructure 
adopting a unique value of $\delta_{c}$ for all clusters would be almost dominated by spurious detections if 
$\delta_{g,95}$ was adopted. The spurious detections are lower if $\delta_{g,99}$ is used. Studies of substructure 
using individual values of $\delta_{c}$ for each cluster will be almost free of spurious substructure galaxies by 
adopting $\delta_{i,max}$. Nevertheless, the fraction of galaxies in substructure detected will be very low. The mean 
fraction of galaxies in substructure detected using $\delta_{i,99}$ is 14$\%$ higher than that adopted with 
$\delta_{i,99.9}$. In contrast, the spurious detections are only 6$\%$ higher with $\delta_{i,99}$ than with 
$\delta_{i,99.9}$. Based on the previous considerations, we adopt $\delta_{i,99}$ and $\delta_{g,99}$ for the 
determination of the galaxies in substructure.

\subsection{Fraction of Galaxies in substructures}

\begin{figure}
   \centering
   \includegraphics[width=9cm]{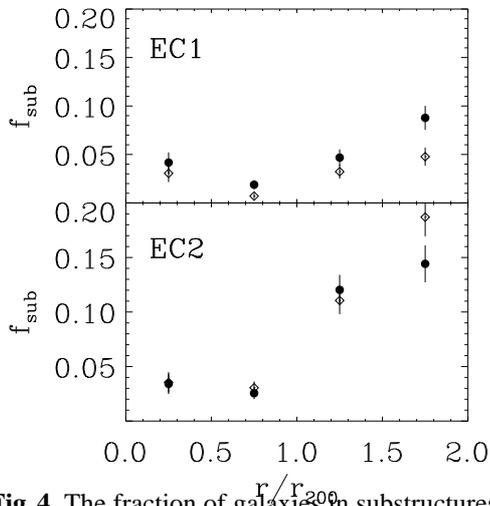}
      \caption{The fraction of galaxies in substructures as a function of cluster radius. The fraction of 
substructure was determined by adopting $\delta_{i,99}$ (full points), $\delta_{g,99}$ (diamonds).
              } 
         \label{f3}
   \end{figure}

\begin{figure}
   \centering
   \includegraphics[width=9cm]{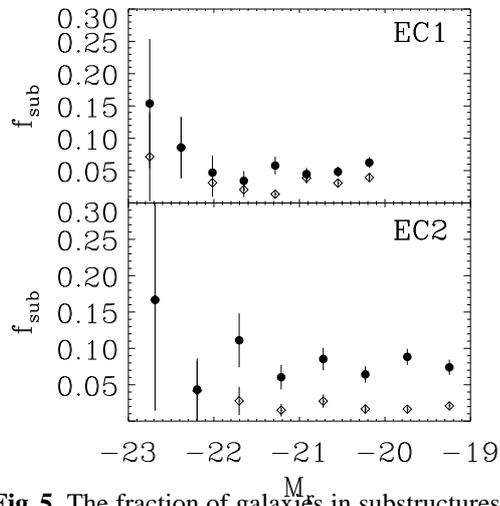}
      \caption{The fraction of galaxies in substructures as a function of galaxy absolute r-band magnitude. The 
fraction of substructure was determined adopting $\delta_{i,99}$ (points), and $\delta_{g,99}$ (diamonds).
              } 
         \label{f4}
   \end{figure}

We computed the fraction of galaxies in substructure as a function of their location in the cluster. 
Figure \ref{f3} shows the fraction of galaxies located in substructures for EC1 and EC2 ensemble clusters. In 
both ensemble clusters, the fraction of galaxies in substructure depends on the $\delta_{c}$ value adopted.  
There is a clear dependence of the fraction of substructure and the location of the galaxies in the cluster. 
In both ensemble clusters, the fraction of substructure in the inner cluster regions ($r<r_{200}$) is smaller than 
in the regions located at $r>r_{200}$.  The EC2 ensemble cluster shows, independently of the $\delta_{c}$ value
adopted, a higher fraction of substructure (especially in the outer cluster region). This shows that the fraction of 
substructure grows as faint galaxies are included in the samples. Note that we have not observed any segregation 
between galaxies in substructures and magnitudes (see Fig. \ref{f4}). This means that the fraction of faint 
galaxies in substructures is not higher than the fraction of bright ones. 

We have investigated the dependence of the previous results with the number of galaxies per cluster. In particular, we have considered those clusters with more than 50 galaxies and we have obtained the fraction of galaxies in substructure as a function of distance to the cluster center and galaxy absolute magnitude. No significant diferences were found for this new cluster sample. The only difference was a small increase in the fraction of galaxies in substructure in the outermost regions of the clusters ($r>r_{200}$). The largest variation was for the EC2 sample. In this case, the fraction of galaxies in substructures peaks at $\approx 25\%$ in the outermost regions.

\begin{figure*}
   \centering
   \includegraphics[width=10cm,angle=90]{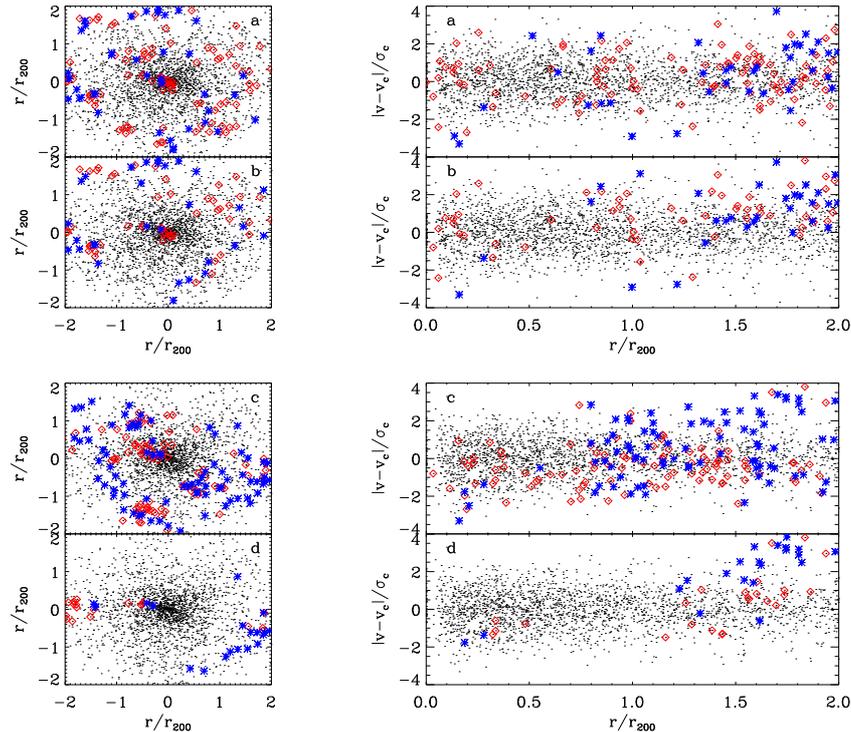}
      \caption{Spatial (left panels) and velocity--radial (right panels) distributions of the galaxies
 outside (points) and inside substructures (diamonds and asterisks). The galaxies in substructures where divided into 
blue (u-r$<2.22$; blue asterisks) and red (u-r$>2.22$; red diamonds). The values of $\delta_{c}$ for identifying galaxies in substructure were: $\delta_{i,99}$ (panels: a and c for EC1 and EC2, respectively), $\delta_{g,99}$ (panels: b and d for EC1 and EC2, respectively).
              } 
         \label{f5}
   \end{figure*}

Figure \ref{f5} shows the spatial and velocity--radial distributions of galaxies inside and outside 
substructures for the two ensemble clusters. Figure \ref{f5} also shows that the galaxies in 
substructure selected with the $\delta_{c}$ values adopted here are mainly located in the outer 
regions of the clusters. We have classified the galaxies into two groups according to their u-r colour. 
Those galaxies showing u-r$<2.22$ were called blue galaxies and those with u-r$>2.22$ were denoted as red. 
According to Strateva et al. (2001), this colour cut separates early- (red) and late-type (blue) galaxies. Note that galaxies 
in substructures located inside $r_{200}$ are mainly red galaxies (about 80$\%$ and 70$\%$ for EC1 and EC2 
respectively). In contrast, the population of red galaxies in substructure located at $r>r_{200}$ is much lower, but still significant  
(about 60$\%$ and 50$\%$ for EC1 and EC2 respectively). The population of blue galaxies inside $r_{200}$ is 
about  20$\%$ for EC1 and 30$\%$ for EC2. Nevertheless, the percentage of these blue galaxies 
located in substructures is very low (less than 5$\%$ in all cases). 

The above percentages of blue and red galaxies can be compared with those obtained from isolated galaxies with similar magnitudes. We have measured these percentages using the sample of isolated galaxies obtained from Allam et al. (2005). Thus, 61$\%$ and 66$\%$ of the isolated galaxies brighter than $M_{r}=-20$ and $M_{r}=-19$ turned to be blue ones, respectively. These percentages are larger than those obtained previously for cluster galaxies. Notice also than the percentage of blue isolated galaxies is also larger than the percentage of blue cluster galaxies located at $r>r_{200}$ and in substructures. This could indicate that galaxies in substructure are not genuine field galaxies. They could be a mixed population of field and cluster galaxies (see the discussion about the dynamical state of galaxies in substructure in Sec. 5.4).

%

%

\section{Discussion}

\subsection{Spurious Substructure Detections}

In the previous sections we have seen that there is a clear difference between the substructure 
presented in the inner ($r<r_{200}$) and outer ($r>r_{200}$) regions of the clusters, 
most of
 the substructure being located in the outermost regions. It is possible that this difference 
between the inner and outer regions of the clusters might not be real, being  due instead to our implementation 
of the DS test. We have investigated this effect using the MC simulations of clusters shown in 
section 3.1. In particular, we have analysed the fraction of substructure as a function of cluster 
radius detected in the MC simulations of clusters built without substructure in them. We need to
 ensure that the large amount of substructure detected in the external regions of the clusters 
($r>r_{200}$) is not spurious. 

\begin{figure}
   \centering
   \includegraphics[width=9cm,angle=0]{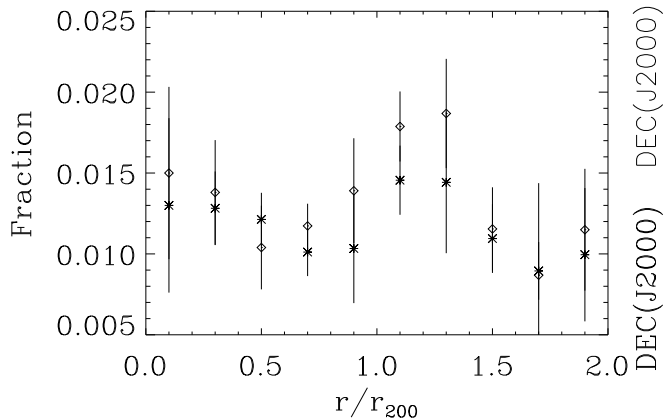}
      \caption{The fraction of substructure as a function of radius in the MC-simulated clusters without 
substructure. The galaxies in substructure were selected using $\delta_{i,99}$ (asterisks) or $\delta_{g,99}$ (diamonds).
              } 
         \label{f6}
   \end{figure}

Figure \ref{f6} shows the fraction of substructure detected in the MC-simulated clusters without 
substructure as function of cluster radius. The substructure was detected by the DS test as in 
the real clusters using a global value of $\delta_{g,99}$ or individual values for the different 
clusters $\delta_{i,99}$ (see section 4). Figure \ref{f6} shows that the fraction of galaxies in
 substructure detected at all radial distances is between 1 and 2 per cent. There is no excess of galaxies 
detected in substructure in the outer regions of the cluster. This rules out the possibility that the 
substructure detected in the outermost regions of the clusters might be dominated by spurious detections.

\subsection{The case of Abell 85}

One of our richest clusters is Abell 85, located at z=0.055 with 273 confirmed members brighter than m$_{r}$=17.88. The substructure of Abell 85 has been previously 
studied in the literature (Ramella et al. 2007; Bravo-Alfaro et al. 2009), and there are also 
available X-ray data for this cluster (Durret et al. 2003), making this cluster 
ideal for comparing the substructure obtained by us with other studies.

Bravo-Alfaro et al. (2009) used the DS test to detect the substructure in Abell 85 and 
found five prominent regions of substructure. The first was located near the centre 
of the cluster and was identified as C2. Another substructure was found to the south called SB, 
and two more appear in the south-east region of Abell 85: one along the X-ray filament detected by 
Durret et al. (2003) was labelled F, and the other along an extension of the latter called SE. 
They also detected a final substructure to the west of the cluster (W). Ramella et al. (2007), using a 
different approach, detected three main structures in Abell 85. The two substructures (S$_{1}$ and S$_{2}$) detected by Ramella et al. (2007) are close to the C2 and SB substructures reported by Bravo-Alfaro et al. (2009).

\begin{figure}
   \centering
   \includegraphics[width=9cm,angle=0]{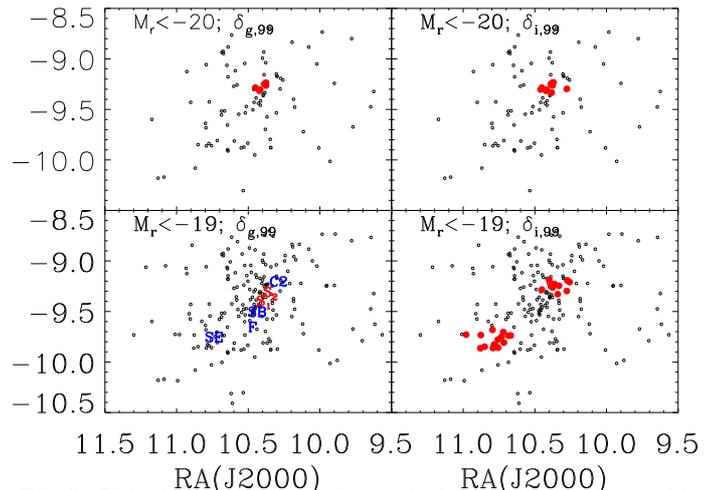}
      \caption{Galaxies in Abell 85 (open circles) and those located in substructure (red full points). 
The top and bottom panels represent the galaxies of Abell 85 for the EC1 and EC2 samples respectively. The 
galaxies in substructure were those with $\delta>\delta_{i,99}$ (right panels) and $\delta>\delta_{g,99}$ (left panels). We have overplotted in the lower-left panel the position of the substructures detected by Bravo-Alfaro et al. (2009; blue color) and Ramella et al. (2007; red color)
              } 
         \label{f9}
   \end{figure}

Figure ~\ref{f9} shows the substructure  in Abell 85 obtained by our method. The substructure galaxies 
were those with $\delta>\delta_{g,99}$ and $\delta>\delta_{i,99}$ for the $M_{r}<-20$ (EC1) and $M_{r}<-19$ 
(EC2) samples. Note that in the EC1 and EC2 samples less  substructure is measured when 
$\delta_{c}=\delta_{g,99}$. Only in the case of EC1 are galaxies in substructure obtained in the 
central region of the cluster. These galaxies are part of the C2 and S$_{2}$ substructure detected by Bravo-Alfaro et al. (2009) and Ramella et al. (2007), respectively. The substructure measured by this method should thus be taken as a lower 
limit of the real value. The situation is better when the galaxies in substructure are determined 
by $\delta_{i,99}$. In this case, for the EC2 sample we obtain two groups of galaxies in substructures 
that correspond to the main groups of galaxies in substructure (C2 and SE) identified by \cite{bravoalfaro09} 
in this cluster. That we have not identified the other groups proposed by these authors 
 might be related to the different galaxy population studied. Their observations are $\sim 2$ magnitudes 
deeper than our data. Notice that we have not identified any galaxy from the substructure labeled F in Bravo-Alfaro et al. (2009). Indeed, at the position of this group of galaxies there are no objects in our data (compare Fig. 4 from Bravo-Alfaro et al. 2009 and our Fig. 7). This could indicate that the galaxies of this substructure are fainter than $M_{r}=-19$ and they are not present in our galaxy sample. As we have seen in previous sections, the completeness of the observations is important
 for the determination of the number of galaxies in substructure. Indeed, variations in the abundance of dwarf galaxies within substructures have been previously observedas, for example, in the Hercules cluster (S\'anchez-Janssen et al. 2005).

We would like also to note that no galaxies in substructure were obtained in Abell 85 for 
the case of EC2 with $\delta<\delta_{g,99}$. This might imply the variation of the $\delta$ values from 
one cluster to another. Some clusters might have galaxies with larger values of $\delta$ that could bias 
$\delta_{g,99}$ to larger values. This is what happened for the EC2 sample; the value of $\delta_{g,99}$ 
is so large that clusters like Abell 85 do not have galaxies in substructure according to this criterion. 
This means that the galaxies in substructure selected in this way would be those with the highest values of 
$\delta$ and should be considered as a lower limit of galaxies in substructures.

\subsection{Luminosity segregation}

The KS test revealed that radial and velocity distributions of galaxies brighter and fainter than 
$M_{r}=-22.0$ were statistically different (at $99\%$ C.L.). This luminosity segregation was also observed in previous 
studies (see Biviano et al. 2002). In most of our clusters, galaxies brighter than $M_{r}=-22.0$ are the 
first rank galaxies of the cluster (84$\%$ and 59$\%$ for EC1 and EC2 respectively). Those bright galaxies 
are mostly located outside substructures. Only $\approx 3-5\%$ (at EC1 and EC2) are located in substructures.
 Moreover, those bright galaxies located outside substructures are very clustered and show similar radial
 velocities to those of the mean cluster ($<r/r_{200}> \approx 0.2$ and $<(v-v_{c})/\sigma_{c}> \approx 0.5$ 
for EC1 and EC2). Nevertheless, those galaxies with $M_{r}<-22.0$ located in substructures are located in 
the outermost regions of the clusters ($<r/r_{200}>=1.05-1.42$ for EC1 and EC2 respectively).

The previous findings could indicate that those galaxies with $M_{r}<-22.0$ and not in substructures would be 
mostly located at the bottom of the cluster potential. In contrast, the small fraction of those bright galaxies 
located in substructures would be falling into the clusters associated with galaxy groups evolved in a merging process. 
The origin of these bright galaxies ($M_{r}<-22.0$) could be due to such accretion or merger processes (Governato 
et al. 2001; de Lucia et al. 2007).

\subsection{Dynamics of galaxies in substructure}

\begin{figure}
   \centering
   \includegraphics[width=9cm]{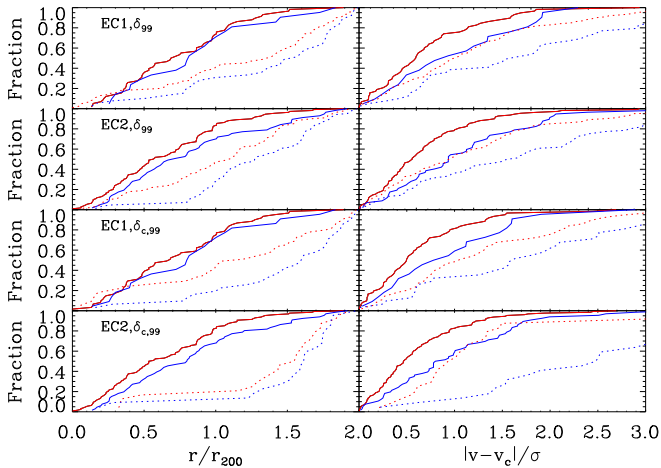}
      \caption{Spatial (left panels) and velocity--radial (right panels) distributions of the galaxies outside
 (full lines) and inside substructures (dotted lines). The galaxies where divided into blue (u-r$<2.22$; blue lines) 
and red (u-r$>2.22$; red lines).
              } 
         \label{f7}
   \end{figure}

Figure~\ref{f7} shows the radial and velocity distributions of galaxies inside and outside substructures. 
We have split the galaxy samples according to their $u-r$ colours. The red and relaxed 
galaxies are always located closer to the cluster centre than the blue and relaxed galaxy sample. Red 
galaxies outside substructures are always cooler than blue and relaxed ones. For galaxies in substructures 
it is also true than the red ones are cooler and are located closer to the cluster centre than the blue 
sample. Nevertheless, galaxies outside substructures are basically inside $r_{200}$, in contrast to what 
happened for galaxies in substructures.

In earlier sections we saw that, in accordance with our selection criteria, our substructure galaxies 
are mainly located beyond $r_{200}$. But can we learn something about the dynamics of these galaxies? 
Are the galaxies in substructure recent arrivals to the cluster potential? Are they an infalling 
sub-population of galaxies?. These questions can be answered by studying the shape of the relative velocity histogram of the galaxies located at $r>r_{200}$.  Gill et al. (2005) investigated the dynamics of satellite galaxies in the 
outskirts of galaxy clusters from a series of high resolution N-body simulations. They found that galaxies 
in clusters located at $r>r_{200}$ were formed by two families: infalling galaxies and the so-called 
back-splash galaxy population. The infalling galaxies are entering the cluster 
potential for the first time. In contrast, the back-splash galaxies are located at large cluster distances ($r>r_{200}$) 
but have previously spent time near the cluster centre. This back-splash galaxy population could be significant 
in number - up to 50$\%$ of 
 the galaxy population located in the region $1.4r_{200}<r<2.8r_{200}$ (see Gill et al. 2005).

The infalling and the back-splash populations can be separated by studying the shape of the relative velocities 
of the galaxies in the outskirts of the clusters. At large distances from the cluster centre ($r>1.4r_{200}$) the 
relative velocity of the infalling galaxies is always higher than that of the back-splash 
galaxies (see Gill et al. 2005). Therefore, if the back-splash galaxy population does not exist, then the relative velocity 
histogram should show a Gaussian-shaped peak at relative velocities greater than zero. In contrast, the presence 
of the back-splash population should distort the Gaussian shape of the relative velocity histogram, peaking at 
zero relative velocity (see Gill et al. 2005). 

\begin{figure*}
   \centering
   \includegraphics[width=15cm]{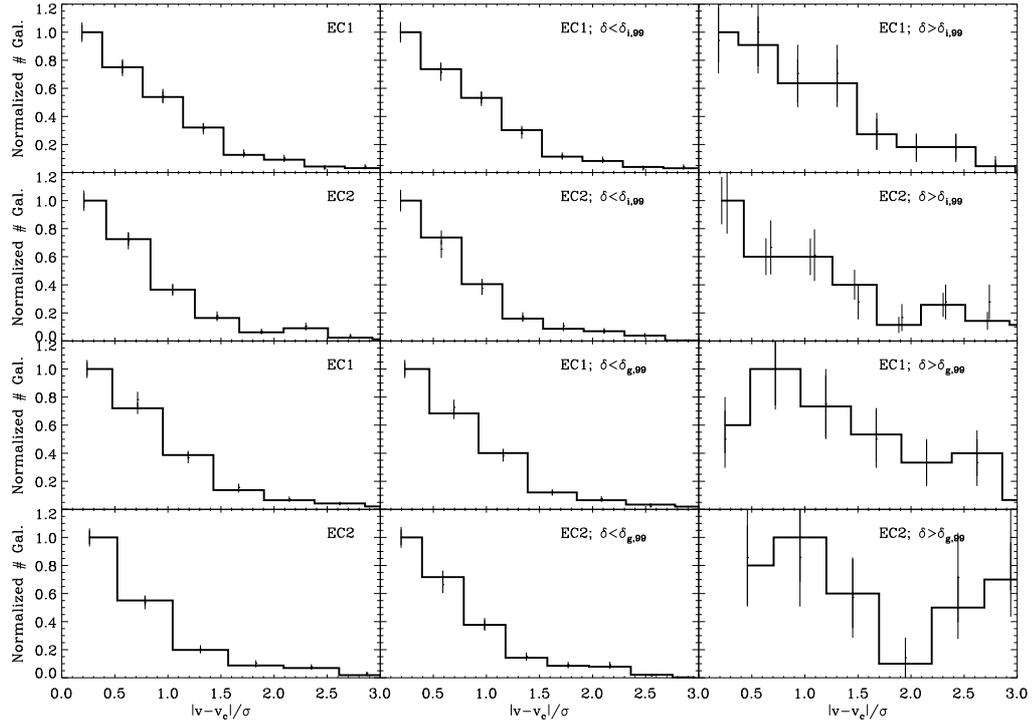}
      \caption{Relative velocity histograms normalized by the velocity dispersion of the clusters for the EC1 
(top panels) and EC2 (bottom panels) clusters. The histograms were normalized by their peak, and correspond to all galaxies 
(left panels), galaxies outside substructures (middle panels), and those inside substructures (right panels). All
 galaxy samples were formed by galaxies located at $1.4r_{200}<r<2.0r_{200}$. The errors correspond to Poisson statistics.
              } 
         \label{f8}
   \end{figure*}

\begin{figure}
   \centering
   \includegraphics[width=9cm]{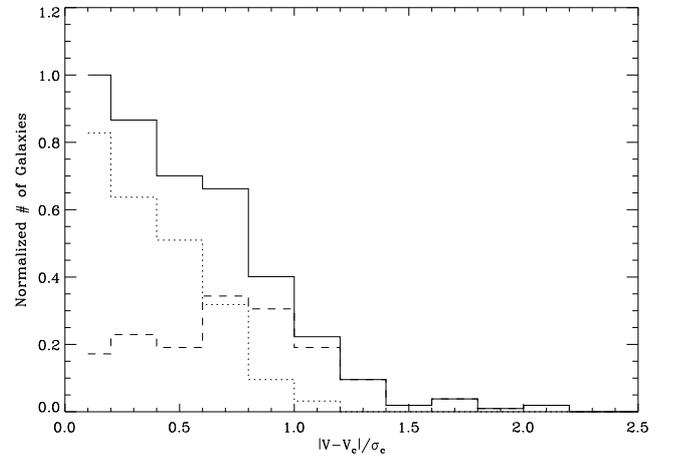}
      \caption{Relative velocity histograms normalized for backsplash (dotted line), infalling (dashed line) and total (full line) galaxy populations from the Gill et al. (2005) and Rines et al. (2005) models of backsplash galaxies. The model showed in this plot has 40$\%$ of the galaxies as an infalling population. 
              } 
         \label{f10}
   \end{figure}

Figure~\ref{f8} shows the relative velocity histograms of all 
the galaxies, and those located inside and outside substructures for our ensemble clusters EC1 and EC2. We considered only those galaxies located at $r>1.4r_{200}$. Figure \ref{f8} indicates that the relative velocity histograms of all the galaxies and those located outside substructures peak at zero velocity. Nevertheless, galaxies in substructures selected by $\delta_{g,99}$ show a peak in the relative velocity histogram different from zero. In order to test the dynamical state of the galaxies in substructures,  we have compared the relative velocities of these galaxies with a mock  velocity distribution of backsplash plus infalling galaxies from fig. 8 of Gill et al. (2005) and Rines et al. (2005). Figure \ref{f10} shows the relative velocity distribution for a model of backsplash and infalling galaxies. We built several models varying the percentage of infalling galaxies. The relative velocity distribution of the models and the observed galaxies in substructures were compared using a KS test. This provides us the percentage of infalling galaxies located in substructure. It should be noticed that the simulations presented by Gill et al. (2005) show results for $1.4r_{200}<r<2.8r_{200}$. Nevertheless, our cluster sample is complete untill 2$r_{200}$. We have investigating the variation of the relative velocity distribution of galaxies in substructure taking into account those galaxies in the range 2-2.8$r_{200}$ for those cluster with galaxies out to 2.8$r_{200}$. No significant difference was found between the relative velocity distributions of galaxies in substructures located in the radius ranges 1.4-2.0$r_{200}$ and 1.4-2.8$r_{200}$.

The KS test gives that at the 95$\%$ C. L. a pure backsplash galaxy population can be ruled out in all cases. Thus, the galaxies in substructure are allways a combination of backplash and infalling galaxies. However, galaxies in substructure selected using $\delta_{i,99}$ show relative velocity histograms dominated by backsplash galaxies. The KS test reports that only 10$\%$ for EC1 and 40$\%$ for EC2 could be infalling galaxies. In contrast, galaxies in substructures selected by $\delta_{g,99}$ are dominated by an infall population. In this case, the KS test gives that 60$\%$ of the galaxies in substructures, for EC1 and EC2 samples, could be infalling galaxies.

The backsplash scenario in the outskirts of the clusters was also studied in the past by Rines et al. (2005) and Pimbblet et al. (2006). Similar to the result presented here, they  found that the outermost galaxy population in clusters is neither a pure backsplash population or purely infalling for the first time. They concluded that local galaxy density is a fundamental parameter for the transformation of galaxies in clusters.

 We have analysed the $u-r$ colours of the galaxies in substructure selected with $\delta_{g,99}$ and located at 1.4$<r/r_{200}<2.0$. The 
blue galaxies ($u-r<2.22$) represent 45$\%$ and 50$\%$ of this  population of galaxies  for EC1 and EC2, respectively. As shown before, $\sim 40\%$ of these galaxies in substructure could be backsplash objects. Assuming, that this percentage of objects would be red galaxies, then the percentage of blue galaxies of the infall populations would be similar to the percentage of blue isolated galaxies (see Sec. 4.1). This means that we can not rule out the hypothesis that the infalling galaxies would be pure field ones, as suggested by recent numerical simulations of cluster formation (see Berrier et al. 2009). 

%

%

\section{Conclusions}

We have analysed the presence of substructure in a sample of 88 nearby ($z<0.1$) and isolated galaxy 
clusters. The galaxy sample consists of 6880 galaxies located in similar physical regions for each galaxy 
clusters ($r<2r_{200}$). The substructure was studied using the DS statistical test. 

The percentage of clusters with substructure is strongly sensitive to the galaxy population mapped by the 
different clusters. Thus, 11$\%$ and 33$\%$ of the clusters show substructure in the inner regions 
($r<r_{200}$) when galaxies brighter than M$_{r}=-20$ or $-19$ are considered, respectively. The 
fraction of cluster with substructure in their outskirts ($r>r_{200}$) is much larger, being 55$\%$ 
and 57$\%$ in the two considered cases. No correlation 
between substructure and global cluster properties ($\sigma_{c}$, $f_{b}$ and $\Delta m_{12}$) has been found. 

We have also studied the galaxies located in substructures by the selection of a 
value of $\delta_{c}$ that distinguishes between galaxies inside and outside substructures. The 
value of $\delta_{c}$ was chosen by comparing the $\delta$ values measured for the galaxies in the 
clusters and those values of $\delta$ from the MC simulations used in the normalization of the DS test. 
This comparison was done individually cluster by cluster, or comparing the global 
distribution of $\delta$ values obtained from stacking the data of the individual clusters into an ensemble one.  In order to avoid possible bias problems with different galaxy population traced
 by the clusters, we have also considered two different galaxy populations, one formed by those galaxies
 brighter than $M_{r}=-20$ located in clusters at $z<0.1$ (called EC1) and the other formed by
galaxies brighter than $M_{r}=-19$ in clusters at $z<0.07$ (called EC 2). 

The fraction of galaxies in substructure increases with radius in both ensemble clusters. Most of the 
galaxies belonging to substructures (especially those detected with a global value of $\delta_{c}$) are located at r$>$r$_{200}$. 
The fraction of galaxies detected in substructure is higher in EC2 than in EC1. Nevertheless,
 we have seen no trend between the fraction of galaxies in substructure and their absolute magnitude. 
Galaxies brighter than $M_{r}=-22$ are preferentially located outside substructures. Those located in 
substructures (only a few percent) are located in the outer regions of the clusters, indicating that they could be located in
 galaxy groups or clusters in the process of merging with the cluster itself.

We have also investigated the dynamics of the galaxies selected in substructures. Independent of the method used for selecting galaxies in substructures, they represent a mixed population of backsplash and infalling galaxies. The substructure galaxies 
selected by a global value of $\delta_{c}$ turned to be dominated by an infalling population of galaxies. In contrast, those galaxies selected in substructures using individial values of $\delta_{c}$ for each cluster turned to be dominated by back-splash galaxies. Assuming that all backsplash galaxies located in substructures are red ones, the fraction of blue galaxies of the infalling population is similar to that observed for isolated objects. This indicates that we can not ruled out the hypothesis that the infall population of galaxies located in substructures would be genuine field ones.

\begin{acknowledgements}
We wish to thank useful comments from anonymous referee which have improved this manuscript. We acknowledge financial support by the grant AYA2007-67965-C03-01. Funding for the SDSS and SDSS-II has been provided by the Alfred P. Sloan Foundation, the Participating Institutions, the National Science Foundation, the U.S. Department of Energy, the National Aeronautics and Space Administration, the Japanese Monbukagakusho, the Max Planck Society, and the Higher Education Funding Council for England. The SDSS Web Site is http://www.sdss.org/. The SDSS is managed by the Astrophysical Research Consortium for the Participating Institutions. The Participating Institutions are the American Museum of Natural History, Astrophysical Institute Potsdam, University of Basel, Cambridge University, Case Western Reserve University, University of Chicago, Drexel University, Fermilab, the Institute for Advanced Study, the Japan Participation Group, Johns Hopkins University, the Joint Institute for Nuclear Astrophysics, the Kavli Institute for Particle Astrophysics and Cosmology, the Korean Scientist Group, the Chinese Academy of Sciences (LAMOST), Los Alamos National Laboratory, the Max-Planck-Institute for Astronomy (MPIA), the Max-Planck-Institute for Astrophysics (MPA), New Mexico State University, Ohio State University, University of Pittsburgh, University of Portsmouth, Princeton University, the United States Naval Observatory, and the University of Washington. This research has made use of the NASA/IPAC Extragalactic Database (NED) which is operated by the Jet Propulsion Laboratory, California Institute of Technology, under contract with the National Aeronautics and Space Administration.
\end{acknowledgements}

\end{document}